\documentclass[ALICE,manyauthors,nocleardouble]{cernphprep}

\usepackage{graphicx, subfigure}
\usepackage{amssymb}
\usepackage{amsmath}
\usepackage{lineno}
\usepackage{cite}
\usepackage{mciteplus}
\usepackage{hyperref}
\usepackage{xspace}
\usepackage{color}

\newcommand{\pp}         {pp\xspace}

\newcommand{\ppb}        {\mbox{p--Pb}\xspace}

\newcommand{\pbpb}       {\mbox{Pb--Pb}\xspace}
\newcommand{\rts}        {\ensuremath{\sqrt{s}}\xspace}
\newcommand{\rtsnn}      {\ensuremath{\sqrt{s_\mathrm{NN}}}\xspace}
\newcommand{\gevc}       {~\ensuremath{\mathrm{GeV}\!/c}\xspace}
\newcommand{\tev}        {~\ensuremath{\mathrm{TeV}}\xspace}
\newcommand{\pt}         {\ensuremath{p_\mathrm{T}}\xspace}
\newcommand{\mpt}        {\ensuremath{\langle\pt\rangle}\xspace}
\newcommand{\nch}        {\ensuremath{N_\mathrm{ch}}\xspace}

\newcommand{\etalab}     {\ensuremath{\eta_\mathrm{lab}}\xspace}

\newcommand{\dndetalab}  {\ensuremath{\mathrm{d}\nch/\mathrm{d}\etalab}\xspace}

\newcommand{\nacc}       {\ensuremath{n_\mathrm{acc}}\xspace}

\newcommand{\mnch}      {\ensuremath{\langle\nch\rangle}\xspace}

\usepackage{multirow}

\begin{document}%

%
\begin{titlepage}
\PHnumber{2013-111}               
\PHdate{2 July 2013}              

\title{Multiplicity dependence of the average transverse momentum \\
in pp, \ppb, and \pbpb collisions at the LHC}

\ShortTitle{Multiplicity dependence of the average transverse momentum}   

\Collaboration{ALICE Collaboration%
         \thanks{See Appendix~\ref{app:collab} for the list of collaboration
                      members}}
\ShortAuthor{ALICE Collaboration} 

\begin{abstract}

The average transverse momentum \mpt versus the
charged-particle multiplicity \nch was measured in 
\ppb collisions at a collision energy per nucleon-nucleon pair \rtsnn = 5.02\tev
and in \pp collisions at collision energies of \rts~=~0.9, 2.76,  
and 7\tev in the kinematic range $0.15<\pt<10.0 \gevc$ and 
$|\eta|<0.3$ with the ALICE apparatus at the LHC.  
These data are compared to results in \pbpb collisions
at \rtsnn = 2.76\tev at similar charged-particle multiplicities.
In \pp and \ppb collisions, a strong increase of \mpt with 
\nch is observed, which is much stronger than that measured in 
\pbpb collisions. 
For \pp collisions, this could be attributed, within 
a model of hadronizing strings, to multiple-parton interactions and to 
a final-state color reconnection mechanism.
The data in \ppb and \pbpb collisions cannot be described by an 
incoherent superposition of nucleon-nucleon collisions and pose a challenge 
to most of the event generators.
\end{abstract}
\end{titlepage}

Measurements of particle production in proton-nucleus collisions 
at the Large Hadron Collider (LHC) energies allow the study of fundamental 
Quantum Chromodynamics (QCD) properties at low parton fractional momentum 
$x$ and high gluon densities; see ~\cite{Salgado:2011wc} for a recent review. 
Additionally, they provide an important reference measurement for studies 
of the properties of the QCD matter created in nucleus-nucleus 
collisions; see ~\cite{Muller:2012zq} for an overview of results at 
the LHC.

The first measurements of charged-particle production in \ppb collisions 
at the LHC at a center-of-mass energy per nucleon-nucleon pair of
\rtsnn = 5.02\tev  \cite{ALICE:2012xs,ALICE:2012mj} exhibited differences 
compared to \pp collisions. These differences were 
mostly confined to low transverse momentum (\pt), leading to a slightly 
smaller average multiplicity per number of participating nucleons in \ppb 
compared to \pp collisions \cite{ALICE:2012xs}, while above a few\gevc the 
\pt  spectrum in \ppb collisions exhibits binary collision scaling 
\cite{ALICE:2012mj}.
The measurements of particle correlations in azimuth and pseudorapidity 
\cite{CMS:2012qk,Abelev:2012ola,Aad:2012gla,Aad:2013fja,Chatrchyan:2013nka} 
have raised the question whether collective effects in \ppb collisions, 
as modeled for example in hydrodynamical approaches \cite{Bozek:2012gr,Shuryak:2013ke}, 
are the origin of the observed correlations.
Initial state effects, such as gluon saturation described by color glass 
condensate (CGC) models \cite{Dusling:2012wy,Dusling:2013oia}, 
reproduce the elliptic flow component, but the triangular flow 
remains a challenge within such models.

It remains questionable if the small system size created in pp or \ppb 
collisions could exhibit collective, fluid-like, features due to early 
thermalization, as observed in \pbpb collisions \cite{Bzdak:2013zma}. 
A meaningful way to address this issue is to investigate production 
mechanisms, correlations, and event shapes as a function of the particle 
multiplicity. 
Such studies were recently performed in pp collisions at the LHC, e.g. 
the ALICE measurements of 
two-pion Bose-Einstein correlations \cite{Aamodt:2011kd},
event sphericity \cite{Abelev:2012sk},
J/$\psi$ meson production \cite{Abelev:2012rz}, 
and anti-baryon to baryon ratios \cite{Abbas:2013rua},
or the measurements by CMS of long-range angular correlations \cite{Khachatryan:2010gv}
and of $\pi$, $K$, and p production \cite{Chatrchyan:2012qb}.

The first moment, \mpt, of the charged-particle transverse momentum spectrum
and its correlation with the charged-particle multiplicity \nch, first observed
at the Sp$\bar{\mathrm p}$S collider \cite{Arnison:1982ed},
carries information about the underlying particle production mechanism.
This has been studied by many experiments at hadron colliders in 
pp($\bar{\mathrm{p}}$) covering collision energies from \rts  =~31~GeV up to 
7~TeV~\cite{ABCDHW, Albajar:1989an, PhysRevLett.60.1622,Adams:2006xb,Aaltonen:2009ne,Aamodt:2010my,Khachatryan:2010nk, Aad:2010ac}. 
All experiments observed an increase of \mpt with \nch in the central rapidity 
region, a feature which could be reproduced in the PYTHIA event generator only 
if a mechanism of hadronization including color correlations (reconnections) 
is considered \cite{Skands:2007zg}.
Although a good description of Tevatron data \cite{Aaltonen:2009ne} was 
achieved within the PYTHIA~8 model \cite{Corke:2010yf}, which also described 
the early LHC data \cite{Corke:2011yy}, full consistency of the data 
description within models is yet to be achieved \cite{Schulz:2011qy}. 
The LHC data highlighted the importance of color 
reconnections \cite{Skands:2010ak};
see also \cite{Schulz:2011qy} and the discussion below.
Data at LHC energies covering a large momentum range starting at 
low \pt provide additional input to these models.

In this letter, we present a measurement of the average transverse momentum 
\mpt versus the charged-particle multiplicity \nch in 
\ppb collisions at a collision energy per nucleon-nucleon pair of
\rtsnn~=~5.02 \tev for primary particles in the kinematic range 
$|\eta| < 0.3$.
These data are compared to results in \pp interactions at collision energies 
of \rts~=~0.9, 2.76, and 7\tev and to results obtained in \pbpb collisions
at \rtsnn~=~2.76\tev. 
The measurements are performed with the ALICE apparatus~\cite{Aamodt:2008zz} 
at the LHC.  
The \pp data were recorded in the years 2009-2011 
and correspond to inelastic collisions, 
details are given in~\cite{Abelev:2013ala}; 
the \pbpb data are from the 2010 run \cite{Abelev:2012eq}.
The \ppb data were recorded during an LHC run of 4 weeks in January 
and February 2013 triggering on non-single-diffractive collisions
\cite{ALICE:2012xs}.
The number of colliding bunches varied between 8 and 288. 
The proton and Pb bunch intensities ranged from 
$1.4\times 10^{10}$ to $1.9\times 10^{10}$ and from 
$0.8\times 10^{10}$ to $1.4\times 10^{10}$ particles, respectively.
The luminosity at the ALICE interaction point was up to
$5\times 10^{27}\mathrm{cm}^{-2}\mathrm{s}^{-1}$ resulting in a hadronic 
interaction rate of $10$~kHz.
The interaction region had an r.m.s. of 6.3~cm along the beam 
direction and about 60~$\mu$m transverse to the beam.
The \ppb minimum-bias events were triggered by requiring a signal in each
of the VZERO detector arrays, VZERO-A located at $2.8<\etalab<5.1$ and VZERO-C 
at $-3.7<\etalab<-1.7$, both covering full azimuth. 
The pseudorapidity of a charged particle in the detector reference-frame 
\etalab is defined as
\begin{math} 
\etalab = -{\rm{ln}}[{\rm{tan}} (\theta/2)]
\end{math}, 
with $\theta$ the polar angle between the beam axis and the charged particle. 
The \pp minimum-bias events were triggered requiring at least a hit in any of 
the VZERO detectors or in the silicon pixel detector covering $|\etalab|<1.4$.

The offline event and track selection is identical to that 
used in the measurement of the charged-particle pseudorapidity density 
\dndetalab \cite{ALICE:2012xs} and the \pt\ spectra in 
\ppb \cite{ALICE:2012mj} and \pbpb \cite{Abelev:2012eq} collisions 
with ALICE.
In total, 106 million events for \ppb collisions, 7, 65, 
and 150 millions for \pp collisions at \rts~=~0.9, 2.76, and 7\tev,
respectively, and 15 millions for \pbpb collisions satisfy the trigger and 
offline event-selection criteria. 
Primary charged particles are defined as all prompt particles produced
in the collision, including all decay products, except those from weak
decays of strange hadrons. 
The efficiency and purity of the primary charged-particle selection are 
estimated from a Monte Carlo simulation using DPMJET~\cite{Roesler:2000he} 
as an event generator with particle transport through the ALICE detector 
using GEANT3~\cite{GEANT3}. 

Due to the asymmetric beam energies for the proton and lead beam, 
the nucleon-nucleon center-of-mass system is moving in the laboratory frame 
with a rapidity of $y_{\mathrm{NN}}=-0.465$; the proton beam has negative 
rapidity. 
In order to ensure good detector acceptance around midrapidity,
tracks are selected for this analysis in the pseudorapidity interval
$|\eta|<0.3$ in the nucleon-nucleon center-of-mass system.
In the absence of information on the particle mass, the particle 
rapidity is unknown.
Therefore, we calculate $\eta = \etalab - y_{\mathrm{NN}}$, 
an approximation which is only accurate for massless particles or 
relativistic particles.
The spectra are corrected based on our knowledge of the pion, kaon, and 
proton yields measured by ALICE \cite{Abelev:2013haa}.
The correction is below 2\% for $\pt<0.5 \gevc$
and below 1\% for $\pt\ge 0.5 \gevc$.
The average transverse momentum \mpt is then calculated from
the corrected spectra as the arithmetic mean in the kinematic range 
$0.15<\pt<10.0 \gevc$ and $|\eta| < 0.3$.
The number of accepted charged particles \nacc is the 
sum of all reconstructed charged particles in the same kinematic range.
To extract the correlation between \mpt\ and the number of primary charged 
particles \nch, counting, for \nch, all particles down to \pt = 0, 
a reweighting procedure is applied to account for the experimental resolution 
in the measured event multiplicity as described in \cite{Aamodt:2010my}. 
This method employs a normalized response matrix from Monte Carlo simulations 
which contains the probability that an event with multiplicity \nch is 
reconstructed with multiplicity \nacc. 

\begin{table}[htb] \centering
\caption{Relative systematic uncertainties on \mpt in \pp, \ppb, and \pbpb 
collisions for $|\eta|<0.3$ and $0.15<\pt<10.0 \gevc$. The quoted ranges
reflect the \nch dependence and, for \pp collisions, also some energy 
dependence.}
\label{systErr}
\begin{tabular}{llll}
\hline\hline
Source                &   \pp      & \ppb & \pbpb \\ \hline
Track selection       & 0.5--1.8\%  & 0.8--1.0\% & 1.1--1.2\%\\
Particle composition  & 0.2--0.4\%  & 0.7--0.8\% & 0.2--0.3\% \\
Tracking efficiency   & 0.1\%      & 0.2\%     & 0.1\%\\
Monte Carlo generator & $\le$0.2\% & 0.1--0.2\% & 0.2\%\\
Reweighting procedure & 2.3--4.1\%  & 1.3--1.8\% & 0.5--1.2\% \\ \hline
Total                 & 2.4--4.5\%  & 1.8--2.2\% & 1.2--3.0\% \\ 
\hline \hline
\end{tabular} \end{table}

The systematic uncertainties of the charged-particle spectrum 
are evaluated in a similar way as in previous analyses of 
\pp \cite{Aamodt:2010my}, \pbpb \cite{Abelev:2012eq}, and 
\ppb \cite{ALICE:2012mj} data and are propagated to \mpt.
The main contributions and the total uncertainties are listed in
Table~\ref{systErr}.
Other contributions investigated are material budget, trigger and event 
selection, and secondary particles from weak decays.
The uncertainty from each of these contributions is below 0.1\%, except the
trigger and event selection, which amounts to 0.35\% for $\nch = 1$.
For \ppb collisions, the effect of the particle composition on the 
uncertainty from acceptance due to the shift in rapidity is included in
Table~\ref{systErr}.
A comparison of the present measurement was performed for the centrality 
classes and the \pt range ($0.3<\pt<2 \gevc$) of the data on pions, kaons and 
protons \cite{Abelev:2013haa}. The agreement is within 0.5\%, well within the 
estimated uncertainty quoted above.
In \pbpb collisions, an additional source of uncertainty at low \nch 
is electromagnetic (EM) processes.
A correction of \mpt of 2.7\% for $\nch = 1$ and less than 1\% for 
$\nch > 5$ was estimated based on a comparison to events in the centrality 
range 0-90\%, where EM events are efficiently rejected \cite{Abelev:2013qoq}.
A conservative systematic uncertainty equal to the correction was assigned 
to this correction and is included in the total uncertainty listed in 
Table~\ref{systErr}.

The uncertainty from the reweighting method is extracted based on the 
Monte Carlo events.
The reweighting procedure is performed using a response matrix
generated with a second event generator and the outcome distribution 
\mpt (\nch) is compared with the initial distribution.
For \pp collisions, PYTHIA6 (Perugia0)~\cite{Skands:2010ak}, 
PYTHIA8~\cite{Sjostrand:2007gs} and PHOJET~\cite{Engel:1995sb} 
event generators are used, while for \ppb and \pbpb collisions we employ
the DPMJET~\cite{Roesler:2000he} and HIJING~\cite{Wang:1991hta} event 
generators.
This uncertainty dominates the overall uncertainty at low \nch, and, 
in \pp collisions, also at large \nch.
An alternative method, based on the integration and extrapolation of \pt
spectra in \nacc bins, gives results well within the systematic uncertainties.

\begin{table}[htb]
\centering
\caption{Characteristics of \pp, \ppb, and \pbpb collisions for events 
with at least one charged particle with $\pt>0.15 \gevc$ in $|\eta|<0.3$. 
The average multiplicity \mnch is for $|\eta|<0.3$ and 
extrapolating to $\pt=0$.
The average transverse momentum \mpt is obtained in $|\eta|<0.3$ and 
in the range $0.15<\pt<10.0 \gevc$. The systematic uncertainties are reported;
the statistical uncertainties are negligible. The uncertainties of 
$\langle\nch\rangle$ are from the tracking efficiency.}

\label{tab_mean} 
\begin{tabular}{c c c c} 
\hline\hline
collision system & \rtsnn (TeV)& \mnch & \mpt (GeV/$c$)  \\ \hline
\pp   &   0.9   &   3.14$\pm$0.16  & 0.540$\pm$0.020 \\ 
\pp   &   2.76  &   3.82$\pm$0.19  & 0.584$\pm$0.020 \\ 
\pp   &   7     &   4.42$\pm$0.22  & 0.622$\pm$0.021 \\
\ppb  &   5.02  &  11.9$\pm$0.5 & 0.696$\pm$0.024 \\  
\pbpb &   2.76  & 259.9$\pm$5.9 & 0.678$\pm$0.007 \\
\hline\hline
\end{tabular}
\end{table}

The values of \mnch and \mpt for events 
with at least one charged particle with $\pt>0.15 \gevc$ in $|\eta|<0.3$
for \pp, \ppb, and \pbpb collisions 
are presented in Table~\ref{tab_mean}. A small increase in \mpt is observed
in \pp collisions as a function of energy. An increase is seen from \pp to 
\ppb and to minimum bias \pbpb collisions.

\begin{figure}[htb]
\begin{center}
\includegraphics[width=0.52\textwidth]{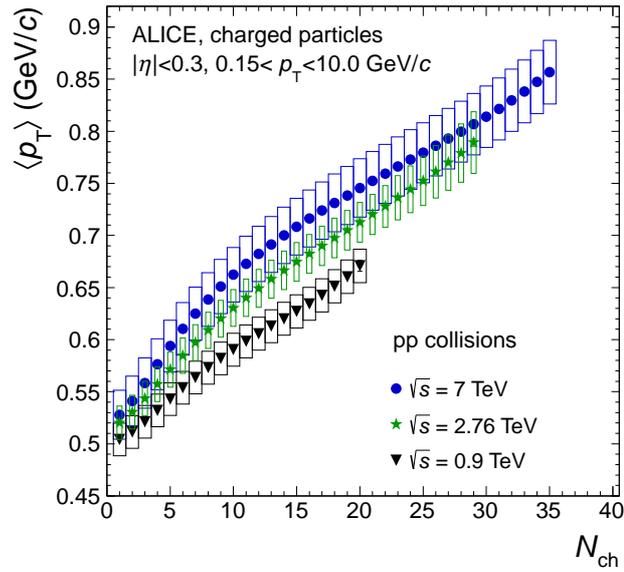}
\end{center}
\caption{Average transverse momentum \mpt in the range $0.15<\pt<10.0 \gevc$ 
as a function of charged-particle multiplicity \nch in \pp collisions at
\rts~=~0.9, 2.76, and 7\tev, for $|\eta|<0.3$. 
The boxes represent the systematic uncertainties on \mpt. 
The statistical errors are negligible.
}
\label{fig_pt0} 
\end{figure}

The average transverse momentum \mpt of charged particles is shown in 
Fig.~\ref{fig_pt0} as a function of the charged-particle multiplicity \nch 
for \pp collisions at \rts~=~0.9, 2.76, and 7\tev.
The multiplicity distributions in \pp collisions 
\cite{Aamodt:2010ft,Aamodt:2010pp} fall off steeply for large \nch.
The present measurement extends up to values of \nch where statistical
errors for \mpt in the corresponding \nacc values are below 5\%.
An increase in \mpt with \nch is observed for all collision energies
and also an increase with the collision energy at fixed values of \nch, which
agrees well with measurements reported by ATLAS \cite{Aad:2010ac,ATLAS-CONF-2010-101} at  \rts~=~0.9 and 7 TeV.
We note a change in slope for all three collision energies at roughly 
the same value of $\nch \approx 10$. 
This change in slope was also observed at Tevatron 
\cite{PhysRevLett.60.1622,Aaltonen:2009ne} and recently at the LHC
\cite{Aad:2010ac,Aamodt:2010my}.

In Monte Carlo event generators, high multiplicity events are produced by
multiple parton interactions. An incoherent superposition of such interactions 
would lead to a constant \mpt at high multiplicities. 
The observed strong correlation of \mpt with \nch has been 
attributed, within PYTHIA models, to color reconnections (CR) between 
hadronizing strings \cite{Skands:2010ak}.
In this mechanism, which can be interpreted as a collective final-state effect, 
strings from independent parton interactions do not hadronize independently, 
but fuse prior to hadronization. This leads to fewer hadrons, but more 
energetic. 
The CR strength is implemented as a probability parameter in the models.
The CR mechanism bears similarity to the mechanism of string fusion 
\cite{Amelin:1993cs} advocated early for nucleus-nucleus collisions.
A model based on Pomeron exchange was shown to fit the pp 
data \cite{Armesto:2008zz}.
A mechanism of collective string hadronization is also used in the EPOS model, 
which was shown recently to describe a wealth LHC data in \pp, \ppb, and
\pbpb collisions \cite{Pierog:2013ria}.

\begin{figure}[htb]
\begin{center}
\includegraphics[width=0.52\textwidth]{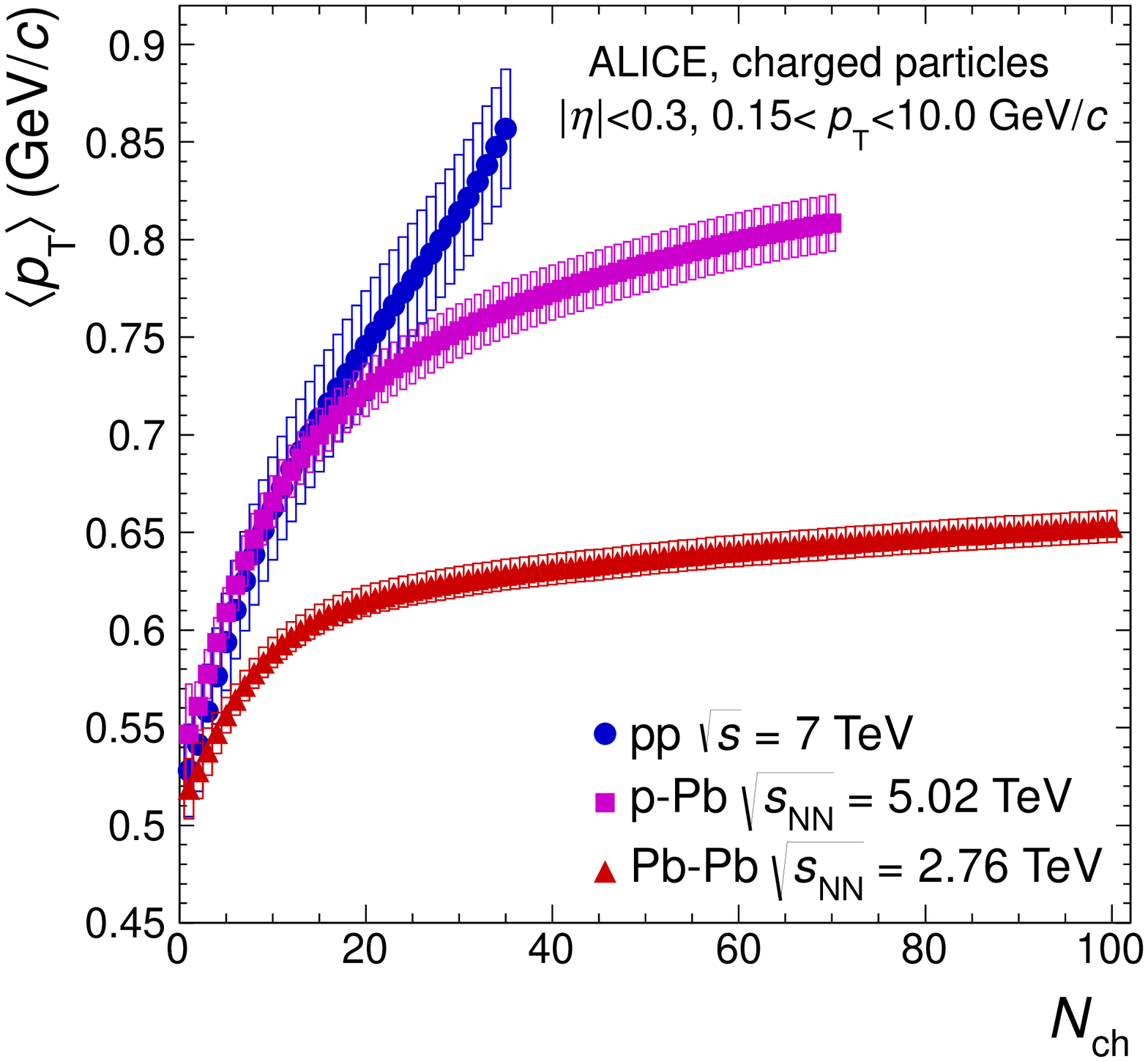}
\end{center}
\caption{Average transverse momentum \mpt versus charged-particle
multiplicity \nch in \pp, \ppb, and \pbpb collisions for $|\eta|<0.3$. 
The boxes represent the systematic uncertainties on \mpt. 
The statistical errors are negligible.}
\label{fig_pt1} 
\end{figure}

Figure~\ref{fig_pt1} shows the average transverse momentum \mpt of 
charged particles versus the charged-particle multiplicity \nch as 
measured in \pp collisions at \rts = 7\tev, in \ppb collisions at 
\rtsnn = 5.02\tev, and in \pbpb collisions at \rtsnn = 2.76\tev.
In \ppb collisions, we observe an increase of \mpt with \nch, with \mpt 
values similar to the values in \pp collisions up to $\nch\approx 14$.
At multiplicities above $\nch\approx 14$, the measured \mpt is lower in 
\ppb collisions than in \pp collisions; the difference is more pronounced 
with increasing \nch.
This difference cannot be attributed to the difference in collision energy, 
as the energy dependence of \mpt is rather weak, see Fig.~\ref{fig_pt0}. 
In contrast, in \pbpb collisions, with increasing \nch, there is only a 
moderate increase in \mpt up to high charged-particle multiplicity with 
a maximum value of $\mpt=0.685\pm 0.016$ (syst.)\gevc, which is substantially 
lower than the maximum value in \pp. 
For \pp and \ppb, $\nch>14$ corresponds to about 10\% and 50\%
of the cross section for events with at least one charged particle with 
$\pt>0.15 \gevc$ in $|\eta|<0.3$, respectively, while for \pbpb collisions 
this fraction is about 82\%; $\nch>40$ corresponds to the upper 
1\% of the cross section in \ppb and to about 70\% most central \pbpb 
collisions.
This illustrates that the same \nch value corresponds to a very 
different collision regime in the three systems.

In \pbpb collisions, substantial rescattering of constituents are thought to 
lead to a redistribution of the particle spectrum where most particles are 
part of a locally thermalized medium exhibiting collective, hydrodynamic-type,
behavior. 
The moderate increase of \mpt seen in \pbpb collisions (in Fig.~\ref{fig_pt1},
for $\nch\gtrsim 10$) is thus usually attributed to collective flow 
\cite{Abelev:2013vea}. 
The \ppb data exhibit features of both \pp and \pbpb collisions, at low and 
high multiplicities, respectively. However, the saturation trend of \mpt versus
\nch is less pronounced in \ppb than in \pbpb collisions and leads to a much 
higher value of \mpt at high multiplicities than in \pbpb. 
An increase in \mpt of a few percent is expected in \pbpb from \rtsnn=2.76 TeV 
to 5 TeV, but it appears unlikely that the \ppb \mpt values will match those
in \pbpb at the same energy.
While the \ppb data cannot exclude collective hydrodynamic-type effects 
for high-multiplicity events, it is clear that such a conclusion requires
stronger evidence.
The features seen in Fig.~\ref{fig_pt1} do not depend on the kinematic 
selection; similar trends are found for $|\eta|<0.8$  ($|\etalab|<0.8$, 
for \ppb collisions) or for $\pt>0.5 \gevc$.

\begin{figure}[t]
\begin{center}
\includegraphics[width=0.52\textwidth]{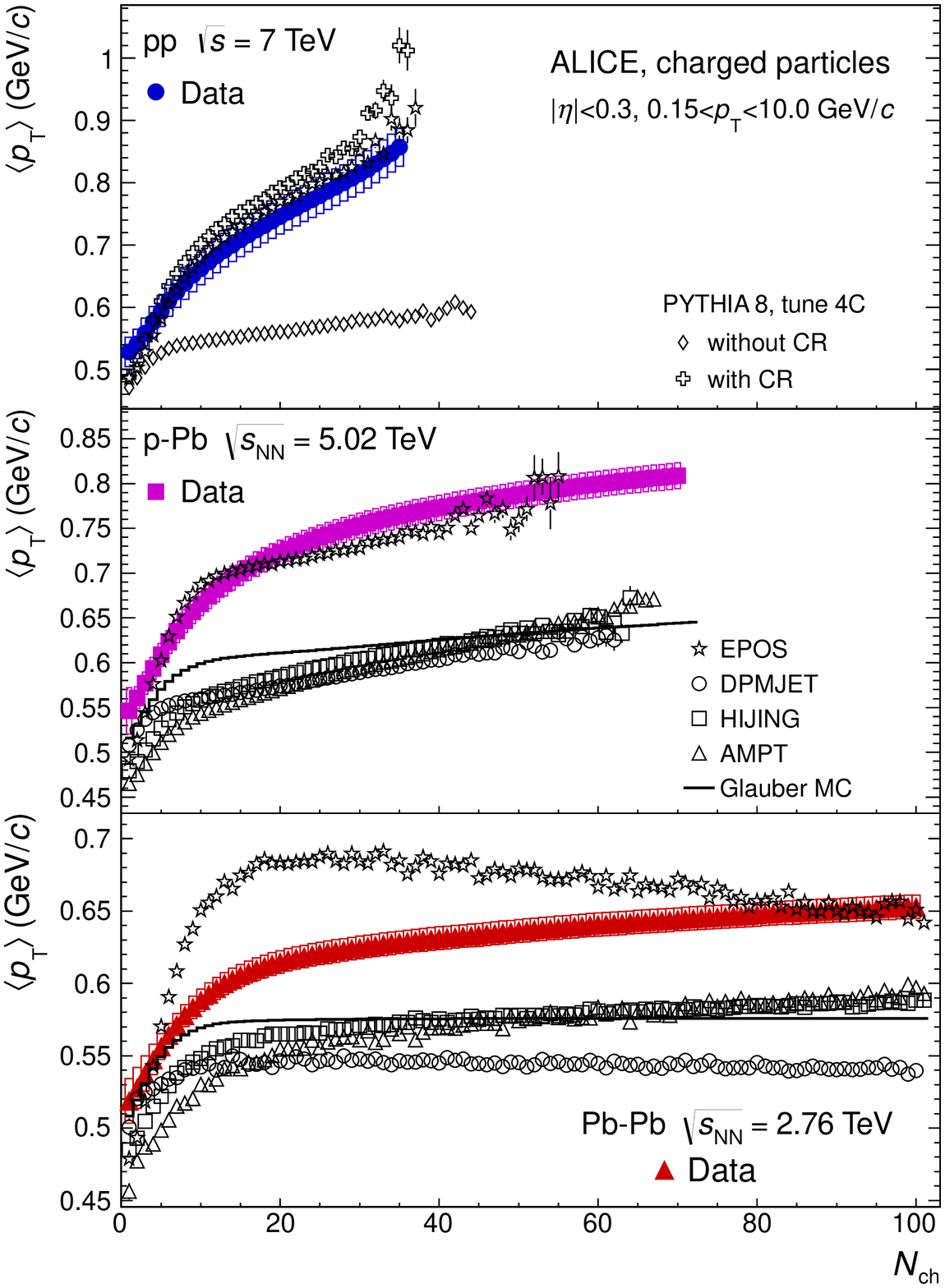}
\end{center}
\caption{Average transverse momentum \mpt as a function of charged-particle 
multiplicity \nch measured in \pp (upper panel), \ppb (middle panel), 
and \pbpb (lower panel) collisions in comparison to model calculations.
The data are compared to calculations with the DPMJET, HIJING, 
AMPT, and EPOS Monte Carlo event generators.
For \pp collisions, calculations with PYTHIA~8 \cite{Sjostrand:2007gs} 
with tune 4C are shown with and without the color reconnection (CR) mechanism.
The lines show calculations in a Glauber Monte Carlo approach (see text).
}
\label{fig_mc} 
\end{figure}

Figure~\ref{fig_mc} shows a comparison of the data to model predictions for 
\mpt versus \nch in \pp collisions at \rts~=~7\tev, \ppb collisions at 
\rtsnn = 5.02\tev and \pbpb collisions at \rtsnn = 2.76\tev.
For \pp collisions, calculations using PYTHIA~8 with tune 4C are shown with 
and without the CR mechanism. As shown earlier 
\cite{Aaltonen:2009ne,Aad:2010ac}, the model only gives a fair description 
of the data when the CR mechanism is included.
Qualitatively, the difference between \ppb and \pbpb collisions seen in 
Fig.~\ref{fig_pt1} is similar to the difference seen in \pp collisions 
between the cases with CR and without CR.
The predictions using the EPOS model (1.99, v3400) describe the data 
well, as expected, given the recent tuning based on the LHC data 
\cite{Pierog:2013ria}. 
In this model collective effects are introduced via parametrizations, for 
the sake of computation time; a full hydrodynamics treatment is available 
in other versions of this model, see \cite{Pierog:2013ria}.
In \ppb collisions, none of the three models, 
DPMJET ~\cite{Roesler:2000he} (v3.0),
HIJING~\cite{Wang:1991hta} (v1.383),
or
AMPT \cite{Lin:2004en} (v2.25, with the string melting option),
describes the data. 
These models predict values of \mpt significantly below the \ppb data. 
The predictions of the EPOS model describe the magnitude of the data but 
show a different trend than data at moderate multiplicities ($\nch<20$).
In addition to predictions from event generators, results of a  
calculation in a Glauber approach are shown. In this approach, 
\ppb collisions are assumed to be a superposition of independent nucleon-nucleon
collisions, each characterized in terms of measured multiplicity 
distributions in \pp collisions \cite{Aamodt:2010ft,Aamodt:2010pp} 
and the \mpt\ values as a function of \nch for \rts=7\tev shown in 
Fig.~\ref{fig_pt0} (for a similar approach, see \cite{Bzdak:2013lva}).
This calculation (continuous line in Fig.~\ref{fig_mc}) underpredicts the 
data, producing, interestingly, results similar to those of event generators. 
The conclusion that \mpt in \ppb collisions is not a consequence of an 
incoherent superposition of nucleon-nucleon collisions invites an analogy
to the observation that \mpt in \pp collisions cannot be described 
by an incoherent superposition of multiple parton interactions.
Whether initial state effects, as considered for the measurement of the 
nuclear modification factor of charged-particle production \cite{ALICE:2012mj},
or final state effects analogous to the CR mechanism are responsible for 
this observation, remains to be further studied.
In \pbpb collisions, the DPMJET, HIJING, and AMPT models fail to 
describe the data, predicting, as in \ppb collisions, lower values of \mpt 
than the measurement. 
The EPOS model overpredicts the data and shows an opposite trend versus
\nch; note, however, that the present model \cite{Pierog:2013ria} includes
collective flow via parametrizations and not a full hydrodynamic treatment.
Also the Glauber MC model with inputs from \mpt data at $\rts=2.76$\tev and 
the measured multiplicity distribution at $\rts=2.36$\tev \cite{Aamodt:2010ft}
fails to describe the data. 

\begin{figure}[htb]
\begin{center}
\includegraphics[width=0.52\textwidth]{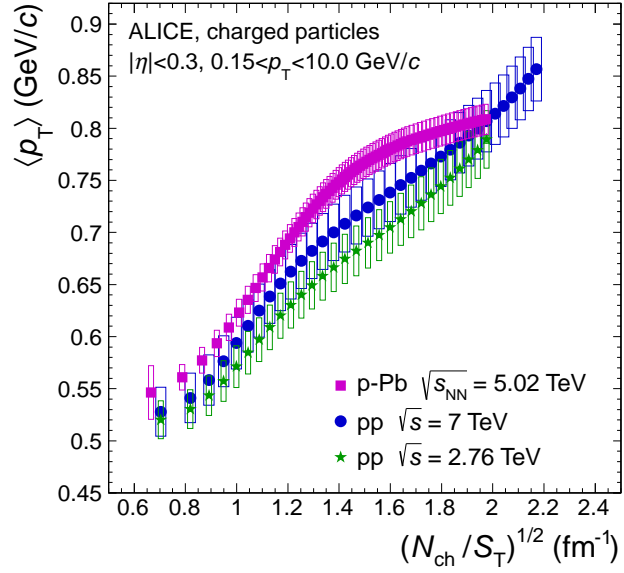}
\end{center}
\caption{Average transverse momentum \mpt 
as a function of the scaled charged-particle multiplicity in \ppb and \pp 
collisions for $|\eta|<0.3$. 
The boxes represent the systematic uncertainties on \mpt. 
The statistical errors are negligible.}
\label{fig_ptx} 
\end{figure}

The data are compared to the geometrical scaling recently proposed 
in \cite{McLerran:2013abc} (and refs. therein) 
within the color-glass condensate model \cite{McLerran:2010ub}.
In this picture, the \mpt is a universal function of the ratio of the 
multiplicity density and the transverse area of the collision,
$S_{\mathrm T}$, calculated within the color-glass model~\cite{Bzdak:2013zma}.
A reasonable agreement was found between this model and CMS data 
\cite{Chatrchyan:2013eya}. 
Employing the parametrizations of $S_{\mathrm T}$ for \pp and \ppb proposed in 
\cite{McLerran:2013abc}, 
the scaling plot in Fig.~\ref{fig_ptx} is obtained.
The ALICE pp data as well as the \ppb data at low and intermediate 
multiplicities are compatible with the proposed scaling. 
As already noted above while discussing Fig.~\ref{fig_pt1} and 
Fig.~\ref{fig_mc}, the behavior of \ppb data at high multiplicities, 
$\nch \gtrsim 14$, shows a departure from the \pp values and cannot be 
described by a binary collision superposition of \pp data. 
The deviation from scaling visible in Fig.~\ref{fig_ptx} for 
$(\nch/S_{\mathrm  T})^{1/2}\gtrsim 1.2$ is related to these observations. 

In summary, we have presented  the average transverse momentum \mpt  
in dependence of the charged-particle multiplicity \nch
measured in \ppb collisions at \rtsnn~=~5.02\tev, in 
\pp collisions at collision energies of \rts~=~0.9, 2.76, and 7\tev and 
in peripheral \pbpb collisions at \rtsnn~=~2.76\tev in the kinematic range 
$0.15<\pt<10.0 \gevc$ and $|\eta| < 0.3$. 
In \pp and \ppb collisions, a strong increase of \mpt with \nch is 
observed, which is understood, in models of \pp collisions, as an effect 
of color reconnections between strings produced in multiple parton interactions.
Whether the same mechanism is at work in \ppb collisions, 
in particular for incoherent proton-nucleon interactions, 
is an open question.
The EPOS model describes the \ppb data assuming collective flow; it remains
to be further studied if initial state effects are compatible with the data.
The \mpt values in \pbpb collisions, instead, indicate a softer 
spectrum and with a much weaker dependence on multiplicity.
These data pose a challenge to most of the existing models and  
are an essential input to improve our understanding of particle 
production as well as the role of initial and final state effects
in these systems. 

%
\newenvironment{acknowledgement}{\relax}{\relax}
\begin{acknowledgement}
\section*{Acknowledgements}

The ALICE collaboration acknowledges the following funding agencies for their support in building and
running the ALICE detector:
 \\
State Committee of Science,  World Federation of Scientists (WFS)
and Swiss Fonds Kidagan, Armenia,
 \\
Conselho Nacional de Desenvolvimento Cient\'{\i}fico e Tecnol\'{o}gico (CNPq), Financiadora de Estudos e Projetos (FINEP),
Funda\c{c}\~{a}o de Amparo \`{a} Pesquisa do Estado de S\~{a}o Paulo (FAPESP);
 \\
National Natural Science Foundation of China (NSFC), the Chinese Ministry of Education (CMOE)
and the Ministry of Science and Technology of China (MSTC);
 \\
Ministry of Education and Youth of the Czech Republic;
 \\
Danish Natural Science Research Council, the Carlsberg Foundation and the Danish National Research Foundation;
 \\
The European Research Council under the European Community's Seventh Framework Programme;
 \\
Helsinki Institute of Physics and the Academy of Finland;
 \\
French CNRS-IN2P3, the `Region Pays de Loire', `Region Alsace', `Region Auvergne' and CEA, France;
 \\
German BMBF and the Helmholtz Association;
\\
General Secretariat for Research and Technology, Ministry of
Development, Greece;
\\
Hungarian OTKA and National Office for Research and Technology (NKTH);
 \\
Department of Atomic Energy and Department of Science and Technology of the Government of India;
 \\
Istituto Nazionale di Fisica Nucleare (INFN) and Centro Fermi -
Museo Storico della Fisica e Centro Studi e Ricerche "Enrico
Fermi", Italy;
 \\
MEXT Grant-in-Aid for Specially Promoted Research, Ja\-pan;
 \\
Joint Institute for Nuclear Research, Dubna;
 \\
National Research Foundation of Korea (NRF);
 \\
CONACYT, DGAPA, M\'{e}xico, ALFA-EC and the EPLANET Program
(European Particle Physics Latin American Network)
 \\
Stichting voor Fundamenteel Onderzoek der Materie (FOM) and the Nederlandse Organisatie voor Wetenschappelijk Onderzoek (NWO), Netherlands;
 \\
Research Council of Norway (NFR);
 \\
Polish Ministry of Science and Higher Education;
 \\
National Authority for Scientific Research - NASR (Autoritatea Na\c{t}ional\u{a} pentru Cercetare \c{S}tiin\c{t}ific\u{a} - ANCS);
 \\
Ministry of Education and Science of Russian Federation, Russian
Academy of Sciences, Russian Federal Agency of Atomic Energy,
Russian Federal Agency for Science and Innovations and The Russian
Foundation for Basic Research;
 \\
Ministry of Education of Slovakia;
 \\
Department of Science and Technology, South Africa;
 \\
CIEMAT, EELA, Ministerio de Econom\'{i}a y Competitividad (MINECO) of Spain, Xunta de Galicia (Conseller\'{\i}a de Educaci\'{o}n),
CEA\-DEN, Cubaenerg\'{\i}a, Cuba, and IAEA (International Atomic Energy Agency);
 \\
Swedish Research Council (VR) and Knut \& Alice Wallenberg
Foundation (KAW);
 \\
Ukraine Ministry of Education and Science;
 \\
United Kingdom Science and Technology Facilities Council (STFC);
 \\
The United States Department of Energy, the United States National
Science Foundation, the State of Texas, and the State of Ohio.
\end{acknowledgement}

\bibliographystyle{href-physrev4}   %
\bibliography{mpt}
\newpage
\appendix
\section{The ALICE Collaboration}
\label{app:collab}



\begingroup
\small
\begin{flushleft}
B.~Abelev\Irefn{org1234}\And
J.~Adam\Irefn{org1274}\And
D.~Adamov\'{a}\Irefn{org1283}\And
A.M.~Adare\Irefn{org1260}\And
M.M.~Aggarwal\Irefn{org1157}\And
G.~Aglieri~Rinella\Irefn{org1192}\And
M.~Agnello\Irefn{org1313}\textsuperscript{,}\Irefn{org1017688}\And
A.G.~Agocs\Irefn{org1143}\And
A.~Agostinelli\Irefn{org1132}\And
Z.~Ahammed\Irefn{org1225}
% ...
\renewcommand\labelenumi{\textsuperscript{\theenumi}~}

\section*{Affiliation notes}
\renewcommand\theenumi{\roman{enumi}}
\begin{Authlist}
\item \Adef{0}Deceased
\item \Adef{idp4108240}{Also at: M.V.Lomonosov Moscow State University, D.V.Skobeltsyn Institute of Nuclear Physics, Moscow, Russia}
\item \Adef{idp4399664}{Also at: University of Belgrade, Faculty of Physics and "Vin\v{c}a" Institute of Nuclear Sciences, Belgrade, Serbia}
\item \Adef{idp5406336}{Also at: Institute of Theoretical Physics, University of Wroclaw, Wroclaw, Poland}
\end{Authlist}

\section*{Collaboration Institutes}
\renewcommand\theenumi{\arabic{enumi}~}
\begin{Authlist}

\item \Idef{org36632}Academy of Scientific Research and Technology (ASRT), Cairo, Egypt
\item \Idef{org1332}A. I. Alikhanyan National Science Laboratory (Yerevan Physics Institute) Foundation, Yerevan, Armenia
\item \Idef{org1279}Benem\'{e}rita Universidad Aut\'{o}noma de Puebla, Puebla, Mexico
\item \Idef{org1220}Bogolyubov Institute for Theoretical Physics, Kiev, Ukraine
\item \Idef{org20959}Bose Institute, Department of Physics and Centre for Astroparticle Physics and Space Science (CAPSS), Kolkata, India
\item \Idef{org1262}Budker Institute for Nuclear Physics, Novosibirsk, Russia
\item \Idef{org1292}California Polytechnic State University, San Luis Obispo, California, United States
\item \Idef{org1329}Central China Normal University, Wuhan, China
\item \Idef{org14939}Centre de Calcul de l'IN2P3, Villeurbanne, France
\item \Idef{org1197}Centro de Aplicaciones Tecnol\'{o}gicas y Desarrollo Nuclear (CEADEN), Havana, Cuba
\item \Idef{org1242}Centro de Investigaciones Energ\'{e}ticas Medioambientales y Tecnol\'{o}gicas (CIEMAT), Madrid, Spain
\item \Idef{org1244}Centro de Investigaci\'{o}n y de Estudios Avanzados (CINVESTAV), Mexico City and M\'{e}rida, Mexico
\item \Idef{org1335}Centro Fermi - Museo Storico della Fisica e Centro Studi e Ricerche ``Enrico Fermi'', Rome, Italy
\item \Idef{org17347}Chicago State University, Chicago, United States
\item \Idef{org1288}Commissariat \`{a} l'Energie Atomique, IRFU, Saclay, France
\item \Idef{org15782}COMSATS Institute of Information Technology (CIIT), Islamabad, Pakistan
\item \Idef{org1294}Departamento de F\'{\i}sica de Part\'{\i}culas and IGFAE, Universidad de Santiago de Compostela, Santiago de Compostela, Spain
\item \Idef{org1106}Department of Physics Aligarh Muslim University, Aligarh, India
\item \Idef{org1121}Department of Physics and Technology, University of Bergen, Bergen, Norway
\item \Idef{org1162}Department of Physics, Ohio State University, Columbus, Ohio, United States
\item \Idef{org1300}Department of Physics, Sejong University, Seoul, South Korea
\item \Idef{org1268}Department of Physics, University of Oslo, Oslo, Norway
\item \Idef{org1315}Dipartimento di Fisica dell'Universit\`{a} and Sezione INFN, Trieste, Italy
\item \Idef{org1145}Dipartimento di Fisica dell'Universit\`{a} and Sezione INFN, Cagliari, Italy
\item \Idef{org1312}Dipartimento di Fisica dell'Universit\`{a} and Sezione INFN, Turin, Italy
\item \Idef{org1285}Dipartimento di Fisica dell'Universit\`{a} `La Sapienza' and Sezione INFN, Rome, Italy
\item \Idef{org1154}Dipartimento di Fisica e Astronomia dell'Universit\`{a} and Sezione INFN, Catania, Italy
\item \Idef{org1132}Dipartimento di Fisica e Astronomia dell'Universit\`{a} and Sezione INFN, Bologna, Italy
\item \Idef{org1270}Dipartimento di Fisica e Astronomia dell'Universit\`{a} and Sezione INFN, Padova, Italy
\item \Idef{org1290}Dipartimento di Fisica `E.R.~Caianiello' dell'Universit\`{a} and Gruppo Collegato INFN, Salerno, Italy
\item \Idef{org1103}Dipartimento di Scienze e Innovazione Tecnologica dell'Universit\`{a} del Piemonte Orientale and Gruppo Collegato INFN, Alessandria, Italy
\item \Idef{org1114}Dipartimento Interateneo di Fisica `M.~Merlin' and Sezione INFN, Bari, Italy
\item \Idef{org1237}Division of Experimental High Energy Physics, University of Lund, Lund, Sweden
\item \Idef{org1192}European Organization for Nuclear Research (CERN), Geneva, Switzerland
\item \Idef{org1227}Fachhochschule K\"{o}ln, K\"{o}ln, Germany
\item \Idef{org1122}Faculty of Engineering, Bergen University College, Bergen, Norway
\item \Idef{org1136}Faculty of Mathematics, Physics and Informatics, Comenius University, Bratislava, Slovakia
\item \Idef{org1274}Faculty of Nuclear Sciences and Physical Engineering, Czech Technical University in Prague, Prague, Czech Republic
\item \Idef{org1229}Faculty of Science, P.J.~\v{S}af\'{a}rik University, Ko\v{s}ice, Slovakia
\item \Idef{org1184}Frankfurt Institute for Advanced Studies, Johann Wolfgang Goethe-Universit\"{a}t Frankfurt, Frankfurt, Germany
\item \Idef{org1215}Gangneung-Wonju National University, Gangneung, South Korea
\item \Idef{org20958}Gauhati University, Department of Physics, Guwahati, India
\item \Idef{org1212}Helsinki Institute of Physics (HIP) and University of Jyv\"{a}skyl\"{a}, Jyv\"{a}skyl\"{a}, Finland
\item \Idef{org1203}Hiroshima University, Hiroshima, Japan
\item \Idef{org1254}Indian Institute of Technology Bombay (IIT), Mumbai, India
\item \Idef{org36378}Indian Institute of Technology Indore, Indore, India (IITI)
\item \Idef{org1266}Institut de Physique Nucl\'{e}aire d'Orsay (IPNO), Universit\'{e} Paris-Sud, CNRS-IN2P3, Orsay, France
\item \Idef{org1277}Institute for High Energy Physics, Protvino, Russia
\item \Idef{org1249}Institute for Nuclear Research, Academy of Sciences, Moscow, Russia
\item \Idef{org1320}Nikhef, National Institute for Subatomic Physics and Institute for Subatomic Physics of Utrecht University, Utrecht, Netherlands
\item \Idef{org1250}Institute for Theoretical and Experimental Physics, Moscow, Russia
\item \Idef{org1230}Institute of Experimental Physics, Slovak Academy of Sciences, Ko\v{s}ice, Slovakia
\item \Idef{org1127}Institute of Physics, Bhubaneswar, India
\item \Idef{org1275}Institute of Physics, Academy of Sciences of the Czech Republic, Prague, Czech Republic
\item \Idef{org1139}Institute of Space Sciences (ISS), Bucharest, Romania
\item \Idef{org27399}Institut f\"{u}r Informatik, Johann Wolfgang Goethe-Universit\"{a}t Frankfurt, Frankfurt, Germany
\item \Idef{org1185}Institut f\"{u}r Kernphysik, Johann Wolfgang Goethe-Universit\"{a}t Frankfurt, Frankfurt, Germany
\item \Idef{org1177}Institut f\"{u}r Kernphysik, Technische Universit\"{a}t Darmstadt, Darmstadt, Germany
\item \Idef{org1256}Institut f\"{u}r Kernphysik, Westf\"{a}lische Wilhelms-Universit\"{a}t M\"{u}nster, M\"{u}nster, Germany
\item \Idef{org1246}Instituto de Ciencias Nucleares, Universidad Nacional Aut\'{o}noma de M\'{e}xico, Mexico City, Mexico
\item \Idef{org1247}Instituto de F\'{\i}sica, Universidad Nacional Aut\'{o}noma de M\'{e}xico, Mexico City, Mexico
\item \Idef{org1308}Institut Pluridisciplinaire Hubert Curien (IPHC), Universit\'{e} de Strasbourg, CNRS-IN2P3, Strasbourg, France
\item \Idef{org1182}Joint Institute for Nuclear Research (JINR), Dubna, Russia
\item \Idef{org1199}Kirchhoff-Institut f\"{u}r Physik, Ruprecht-Karls-Universit\"{a}t Heidelberg, Heidelberg, Germany
\item \Idef{org20954}Korea Institute of Science and Technology Information, Daejeon, South Korea
\item \Idef{org1017642}KTO Karatay University, Konya, Turkey
\item \Idef{org1160}Laboratoire de Physique Corpusculaire (LPC), Clermont Universit\'{e}, Universit\'{e} Blaise Pascal, CNRS--IN2P3, Clermont-Ferrand, France
\item \Idef{org1194}Laboratoire de Physique Subatomique et de Cosmologie (LPSC), Universit\'{e} Joseph Fourier, CNRS-IN2P3, Institut Polytechnique de Grenoble, Grenoble, France
\item \Idef{org1187}Laboratori Nazionali di Frascati, INFN, Frascati, Italy
\item \Idef{org1232}Laboratori Nazionali di Legnaro, INFN, Legnaro, Italy
\item \Idef{org1125}Lawrence Berkeley National Laboratory, Berkeley, California, United States
\item \Idef{org1234}Lawrence Livermore National Laboratory, Livermore, California, United States
\item \Idef{org1251}Moscow Engineering Physics Institute, Moscow, Russia
\item \Idef{org1322}National Centre for Nuclear Studies, Warsaw, Poland
\item \Idef{org1140}National Institute for Physics and Nuclear Engineering, Bucharest, Romania
\item \Idef{org1017626}National Institute of Science Education and Research, Bhubaneswar, India
\item \Idef{org1165}Niels Bohr Institute, University of Copenhagen, Copenhagen, Denmark
\item \Idef{org1109}Nikhef, National Institute for Subatomic Physics, Amsterdam, Netherlands
\item \Idef{org1283}Nuclear Physics Institute, Academy of Sciences of the Czech Republic, \v{R}e\v{z} u Prahy, Czech Republic
\item \Idef{org1264}Oak Ridge National Laboratory, Oak Ridge, Tennessee, United States
\item \Idef{org1189}Petersburg Nuclear Physics Institute, Gatchina, Russia
\item \Idef{org1170}Physics Department, Creighton University, Omaha, Nebraska, United States
\item \Idef{org1157}Physics Department, Panjab University, Chandigarh, India
\item \Idef{org1112}Physics Department, University of Athens, Athens, Greece
\item \Idef{org1152}Physics Department, University of Cape Town and  iThemba LABS, National Research Foundation, Somerset West, South Africa
\item \Idef{org1209}Physics Department, University of Jammu, Jammu, India
\item \Idef{org1207}Physics Department, University of Rajasthan, Jaipur, India
\item \Idef{org1200}Physikalisches Institut, Ruprecht-Karls-Universit\"{a}t Heidelberg, Heidelberg, Germany
\item \Idef{org1017688}Politecnico di Torino, Turin, Italy
\item \Idef{org1325}Purdue University, West Lafayette, Indiana, United States
\item \Idef{org1281}Pusan National University, Pusan, South Korea
\item \Idef{org1176}Research Division and ExtreMe Matter Institute EMMI, GSI Helmholtzzentrum f\"ur Schwerionenforschung, Darmstadt, Germany
\item \Idef{org1334}Rudjer Bo\v{s}kovi\'{c} Institute, Zagreb, Croatia
\item \Idef{org1298}Russian Federal Nuclear Center (VNIIEF), Sarov, Russia
\item \Idef{org1252}Russian Research Centre Kurchatov Institute, Moscow, Russia
\item \Idef{org1224}Saha Institute of Nuclear Physics, Kolkata, India
\item \Idef{org1130}School of Physics and Astronomy, University of Birmingham, Birmingham, United Kingdom
\item \Idef{org1338}Secci\'{o}n F\'{\i}sica, Departamento de Ciencias, Pontificia Universidad Cat\'{o}lica del Per\'{u}, Lima, Peru
\item \Idef{org1155}Sezione INFN, Catania, Italy
\item \Idef{org1313}Sezione INFN, Turin, Italy
\item \Idef{org1271}Sezione INFN, Padova, Italy
\item \Idef{org1133}Sezione INFN, Bologna, Italy
\item \Idef{org1146}Sezione INFN, Cagliari, Italy
\item \Idef{org1316}Sezione INFN, Trieste, Italy
\item \Idef{org1115}Sezione INFN, Bari, Italy
\item \Idef{org1286}Sezione INFN, Rome, Italy
\item \Idef{org36377}Nuclear Physics Group, STFC Daresbury Laboratory, Daresbury, United Kingdom
\item \Idef{org1258}SUBATECH, Ecole des Mines de Nantes, Universit\'{e} de Nantes, CNRS-IN2P3, Nantes, France
\item \Idef{org35706}Suranaree University of Technology, Nakhon Ratchasima, Thailand
\item \Idef{org1304}Technical University of Split FESB, Split, Croatia
\item \Idef{org1017659}Technische Universit\"{a}t M\"{u}nchen, Munich, Germany
\item \Idef{org1168}The Henryk Niewodniczanski Institute of Nuclear Physics, Polish Academy of Sciences, Cracow, Poland
\item \Idef{org17361}The University of Texas at Austin, Physics Department, Austin, TX, United States
\item \Idef{org1173}Universidad Aut\'{o}noma de Sinaloa, Culiac\'{a}n, Mexico
\item \Idef{org1296}Universidade de S\~{a}o Paulo (USP), S\~{a}o Paulo, Brazil
\item \Idef{org1149}Universidade Estadual de Campinas (UNICAMP), Campinas, Brazil
\item \Idef{org1239}Universit\'{e} de Lyon, Universit\'{e} Lyon 1, CNRS/IN2P3, IPN-Lyon, Villeurbanne, France
\item \Idef{org1205}University of Houston, Houston, Texas, United States
\item \Idef{org20371}University of Technology and Austrian Academy of Sciences, Vienna, Austria
\item \Idef{org1222}University of Tennessee, Knoxville, Tennessee, United States
\item \Idef{org1310}University of Tokyo, Tokyo, Japan
\item \Idef{org1318}University of Tsukuba, Tsukuba, Japan
\item \Idef{org21360}Eberhard Karls Universit\"{a}t T\"{u}bingen, T\"{u}bingen, Germany
\item \Idef{org1225}Variable Energy Cyclotron Centre, Kolkata, India
\item \Idef{org1017687}Vestfold University College, Tonsberg, Norway
\item \Idef{org1306}V.~Fock Institute for Physics, St. Petersburg State University, St. Petersburg, Russia
\item \Idef{org1323}Warsaw University of Technology, Warsaw, Poland
\item \Idef{org1179}Wayne State University, Detroit, Michigan, United States
\item \Idef{org1143}Wigner Research Centre for Physics, Hungarian Academy of Sciences, Budapest, Hungary
\item \Idef{org1260}Yale University, New Haven, Connecticut, United States
\item \Idef{org15649}Yildiz Technical University, Istanbul, Turkey
\item \Idef{org1301}Yonsei University, Seoul, South Korea
\item \Idef{org1327}Zentrum f\"{u}r Technologietransfer und Telekommunikation (ZTT), Fachhochschule Worms, Worms, Germany
\end{Authlist}
\endgroup

%

\end{document}